\def\dash{\hbox{--}}
\def\degree{{$^\circ$}}
\def\ts{\thinspace}           

\def\puncspace{\ifmmode\,\else{\ifcat.\C{\if.\C\else%
\if,\C\else\if?\C\else\if:\C\else\if;\C\else\if-\C\else%
\if)\C\else\if/\C\else\if]\C\else\if'\C%
\else\space\fi\fi\fi\fi\fi\fi\fi\fi\fi\fi}%
\else\if\empty\C\else\if\space\C\else\space\fi\fi\fi}\fi}%
\def\SP{\let\\=\empty\futurelet\C\puncspace}

\def\ee#1{\ifmmode {} \times 10^{#1} \else ${} \times 10^{#1}$\fi}
\def\sub#1{\ifmmode _{#1} \else $_{#1}$\fi}
\def\sup#1{\ifmmode ^{#1} \else $^{#1}$\fi}

\def\dash{\hbox{--}}
\def\about{\ifmmode \sim \else {$\sim\,$}\fi}
\def\lta{\ifmmode {\,\mathbin{\lower 3pt\hbox   
    {$\,\rlap{\raise 5pt\hbox{$\char'074$}}\mathchar"7218\,$}}}
    \else {${\mathbin{\lower 3pt\hbox
    {$\rlap{\raise 5pt\hbox{$\char'074$}}\mathchar"7218\,$}}}
    $}\fi}
\def\gta{\ifmmode {\mathbin{\lower 3pt\hbox   
    {$\,\rlap{\raise 5pt\hbox{$\char'076$}}\mathchar"7218\,$}}}
    \else {${\mathbin{\lower 3pt\hbox
    {$\rlap{\raise 5pt\hbox{$\char'076$}}\mathchar"7218\,$}}}
    $}\fi}

\def\ast{\mathchar"2203} \mathcode`*="002A   

\def\degree{{\ifmmode ^\circ \else $^\circ$\fi}}

\def\Hz{{\hbox{Hz}}\SP}

\def\fu#1{\leavevmode\hbox{4U~#1}\SP}

\def\ks#1{\leavevmode\hbox{KS~#1}\SP}

\def\exosat{\leavevmode{\it EXOSAT}\SP}
\def\ginga{\leavevmode{\it Ginga\/}\SP}

\def\mdot{{\ifmmode \dot M \else {$\dot M$}\fi}}
\def\mdote{{\ifmmode \dot M_E \else {$\dot M_E$}\fi}}
\def\mdoti{{\ifmmode \dot M_i \else {$\dot M_i$}\fi}}
\def\msun{{\ifmmode M_\odot \else {$M_{\odot}$}\fi}}
\def\nonrot{{\rm 0}}

\documentstyle{neu}
\input psfig.sty
\begin{document}

\title{%
 THE ORIGIN OF KILOHERTZ QPOs AND 
 IMPLICATIONS FOR NEUTRON STARS}

 \author{Frederick K. LAMB\\ {\it 
University of Illinois at
Urbana-Champaign, Departments of Physics
and Astronomy, 1110 W. Green St., Urbana,
IL 61801, USA, f-lamb@uiuc.edu}\\
 M. Coleman MILLER\\
 {\it University of Chicago, Department of
Astronomy and Astrophysics, 5640 S. Ellis
Ave., Chicago, IL 60637,
miller@bayes.uchicago.edu}
 Dimitrios PSALTIS\\
 {\it Harvard-Smithsonian Center for
Astrophysics, 60 Garden St., Cambridge, MA
02138, USA, dpsaltis@cfa.harvard.edu}
 }

\maketitle

\section*{Abstract}

 One of the most dramatic discoveries made
with the {\em Rossi X-Ray Timing
Explorer\/} is that many accreting neutron
stars in low-mass binary systems produce
strong, remarkably coherent, high-frequency
X-ray brightness oscillations.
 The $\sim$325--1200~Hz 
quasi-periodic oscillations (QPOs) observed
in the accretion-powered emission are
thought to be produced by gas orbiting very
close to the neutron star, whereas the
$\sim$360--600~Hz brightness oscillations
seen during thermonuclear X-ray bursts are
produced by one or two hot spots rotating
with the star and have frequencies equal to
the stellar spin frequency or its first
overtone.
 The oscillations constrain the masses and
radii of these neutron stars, which are
thought to be the progenitors of the
millisecond pulsars. Modeling indicates
that the stars have spin frequencies
$\sim$250--350~Hz and magnetic fields
$\sim$10$^{7}$ -- $5\ee{9}$~G.

\section{Introduction}

The discovery of strong and remarkably
coherent high-frequency X-ray brightness
oscillations in at least sixteen neutron
stars in low-mass binary systems has
provided valuable new information about
these stars, some of which are likely to
become millisecond pulsars. Oscillations
are observed both in the persistent X-ray
emission and during thermonuclear X-ray
bursts (see van der Klis 1997).

The kilohertz quasi-periodic oscillations
(QPOs) observed in the persistent emission
have frequencies in the range 325--1200~Hz,
amplitudes as high as $\sim15$\%, and
quality factors $\nu/\delta\nu$ as high as
$\sim200$. Two kilohertz QPOs are commonly
observed simultaneously in a given source
(see Fig.~1). Although the frequencies of
the two QPOs vary by  hundreds of Hertz, the
frequency separation $\Delta\nu$ between
them appears to be nearly constant in almost
all cases (see van der Klis et al.\
1997 and M\'endez et al.\ 1997).

 \begin{figure}[t] 
 \begin{minipage}[b]{3.3in}
 \centerline{
\psfig{file=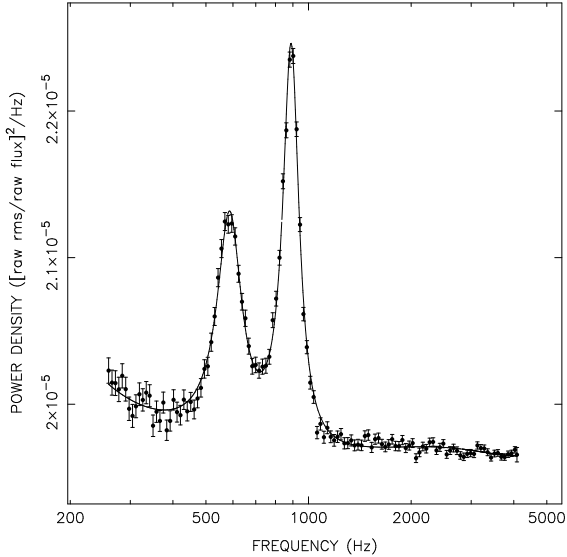,height=7.5truecm}}
 \end{minipage}
 \begin{minipage}[b]{2.3in}
 \caption{Power density spectrum of
Sco~X-1 brightness variations, showing the
two simultaneous kilohertz QPOs that are
characteristic.
 These are two of the weakest kilohertz
QPOs observed, with rms amplitudes
$\sim1$\%.
 The continuum power density is consistent
with that expected from photon counting
noise.
 From van der Klis et al.\
(1997).}
 \end{minipage}
 \end{figure}

The $\sim$250--600~Hz brightness
oscillations observed during type~I X-ray
bursts are different in character from the
QPOs observed in the persistent emission
(see Strohmayer, Zhang, \& Swank 1997).
Only a single oscillation has been observed
during X-ray bursts, and the oscillations
in the tails of bursts appear to be highly
coherent (see, e.g., Smith, Morgan, \&
Bradt 1997), with frequencies that are
always the same for a given source
(comparison of burst oscillations from
\fu{1728$-$34} over about a year shows that
the timescale for any variation in the
oscillation frequency is $\gta 3000$~yr;
Strohmayer 1997). The burst oscillations in
\fu{1728$-$34} and \fu{1702$-$42} (see
Strohmayer, Swank, \& Zhang 1998) have
frequencies that are consistent with the
separation frequencies of their kilohertz
QPO pairs. The burst oscillations in
\fu{1636$-$536} (Zhang et al.\ 1997) and
\ks{1731$-$260} (Smith et al.\ 1997) have
frequencies that are consistent with twice
the separation frequencies of their
kilohertz QPO pairs (Zhang et al.\ 1997;
Wijnands \& van der Klis 1997). The
evidence is compelling that the burst
oscillations are produced by rotation with
the star of one or two nearly identical
emitting spots on the surface (see
Strohmayer et al.\ 1997). The frequencies
of the burst oscillations are therefore the
stellar spin frequency or its first
overtone.

The frequency separation $\Delta\nu$ between
the two kilohertz QPOs observed in the
persistent emission of a given star is
closely equal to the spin frequency of the
star inferred from its burst oscillations
(see Miller, Lamb, \& Psaltis 1998,
hereafter MLP).

\section{Origin of Kilohertz QPOs}

Although other types of models have been
suggested (Klein et~al.\ 1996; Titarchuk \&
Muslimov 1997), the evidence favoring
beat-frequency models of the two kilohertz
QPOs is very strong (see van der Klis
1997), and we therefore focus on these.

The magnetospheric beat-frequency model was
developed to explain the single,
$\sim$15--60~Hz ``horizontal branch 
oscillation'' (HBO) observed in the Z
sources (see Lamb 1991). In this model, the
frequency of the HBO is the difference
between the Keplerian orbital frequency
$\nu_{\rm Km}$ at the main radius where the
stellar magnetic field picks up and
channels gas from the accretion disk onto
the magnetic polar regions and the stellar
spin frequency $\nu_{\rm spin}$. Strohmayer
et~al.\ (1996) applied the magnetospheric
beat-frequency idea to the kilohertz QPO
pairs, interpreting the frequency of the
higher-frequency QPO in a pair as $\nu_{\rm
Km}$ and the frequency of the
lower-frequency QPO as \hbox{$\nu_{\rm Km} -
\nu_{\rm spin}$}. Although it explains
naturally why the frequency separation
between the QPOs in a pair is nearly
constant in most sources and equal to the
burst oscillation frequency or half this
frequency, there are many serious
difficulties with the magnetospheric
beat-frequency interpretation of the
kilohertz QPOs (see MLP).

The most fully developed and successful
model of the kilohertz QPOs is the so-called
sonic-point beat-frequency model, in which
the higher frequency in a QPO pair is the
orbital frequency of gas at the inner edge
of the Keplerian disk flow and the lower
frequency is the difference between this
frequency and the spin frequency of the
neutron star. The sonic-point model was
developed (MLP) specifically to explain the
kilohertz QPO pairs and is based on
previous work (Miller \& Lamb 1996) which
showed that the drag force produced by
radiation from a central star can terminate
a Keplerian disk flow near the star. In the
sonic-point model, some accreting gas
spirals inward in nearly circular Keplerian
orbits until it is close to the neutron
star, where radiation forces or general
relativistic effects cause a sudden
increase in the inward radial velocity,
which becomes supersonic within a small
radial distance (see Fig.~2a). The sharp
increase in the radial velocity is usually
caused by the drag exerted on the accreting
gas by radiation from the star, but may
instead be caused by general relativistic
corrections to Newtonian gravity if the gas
in the Keplerian disk flow reaches the
innermost stable circular orbit without
being significantly affected by radiation.

As the disk flow approaches the sonic point,
the optical depth of the flow in the radial
direction (measured from the stellar
surface) typically falls steeply with
decreasing radius (see Fig.~2b). The change
from optically thick to optically thin disk
flow occurs within a few photon mean free
paths and is somewhat analogous to the
ionization front at the boundary of an
H{\ts\scriptsize II} region, except that
here the photon mean free path increases
because the radiation is removing angular
momentum and the flow is accelerating
inward, causing the density to fall
sharply, whereas in an ionization front the
mean free path is increases because
radiation is removing bound electrons from
atoms and molecules, causing the opacity to
fall sharply. Once the accreting gas is
exposed to the radiation from the star, it
loses its angular momentum to radiation
drag in a radial distance $\delta \lta
0.01\,r$. It then falls inward
supersonically along spiral trajectories
and collides with the neutron star around
its equator, producing an X-ray emitting
equatorial ring.

 \begin{figure*}[t] 
 \label{transition}
 \vglue-1.5truecm
 \hbox{\hglue-0truecm
{\psfig{file=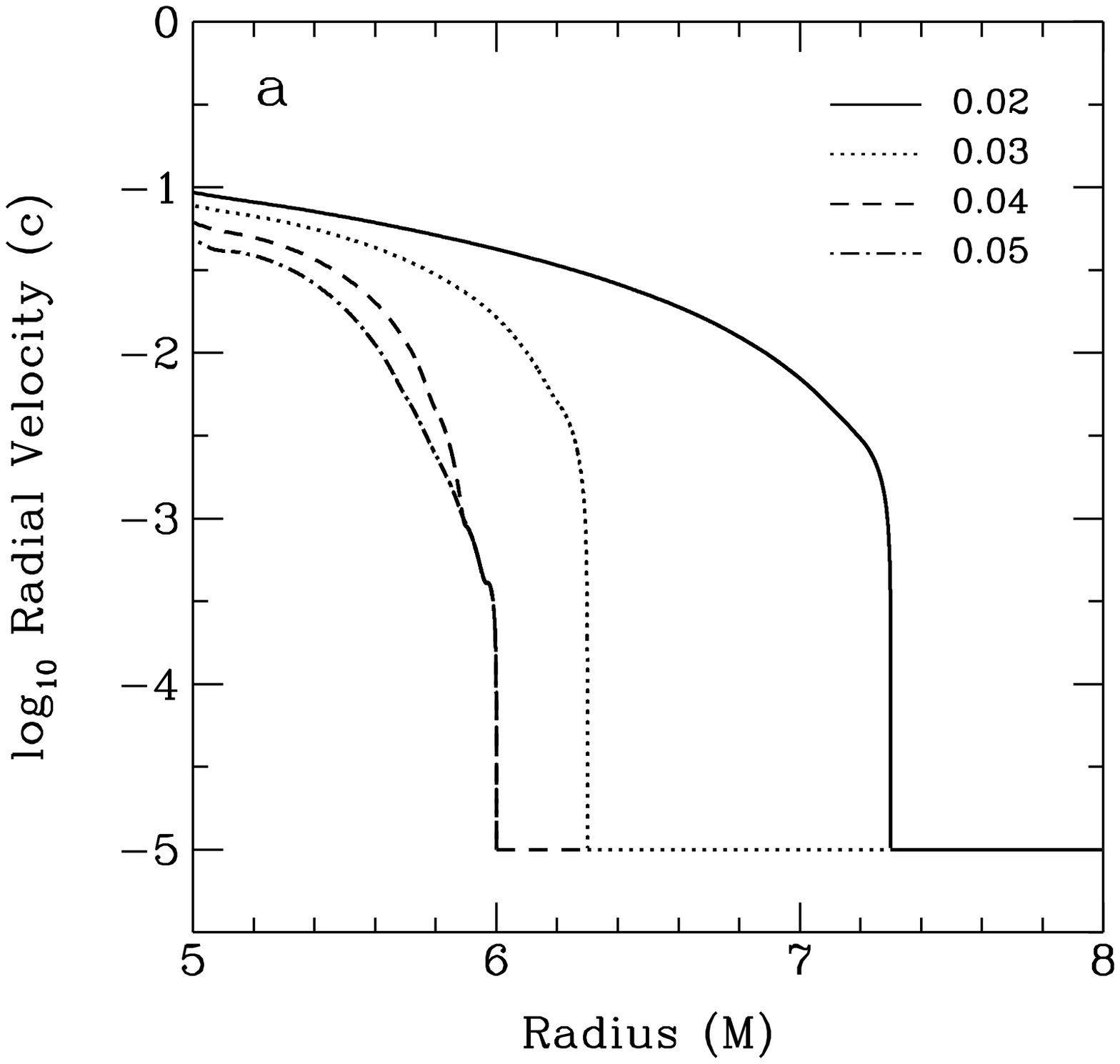,height=3.4truein}}
 \hglue-1.4truecm
{\psfig{file=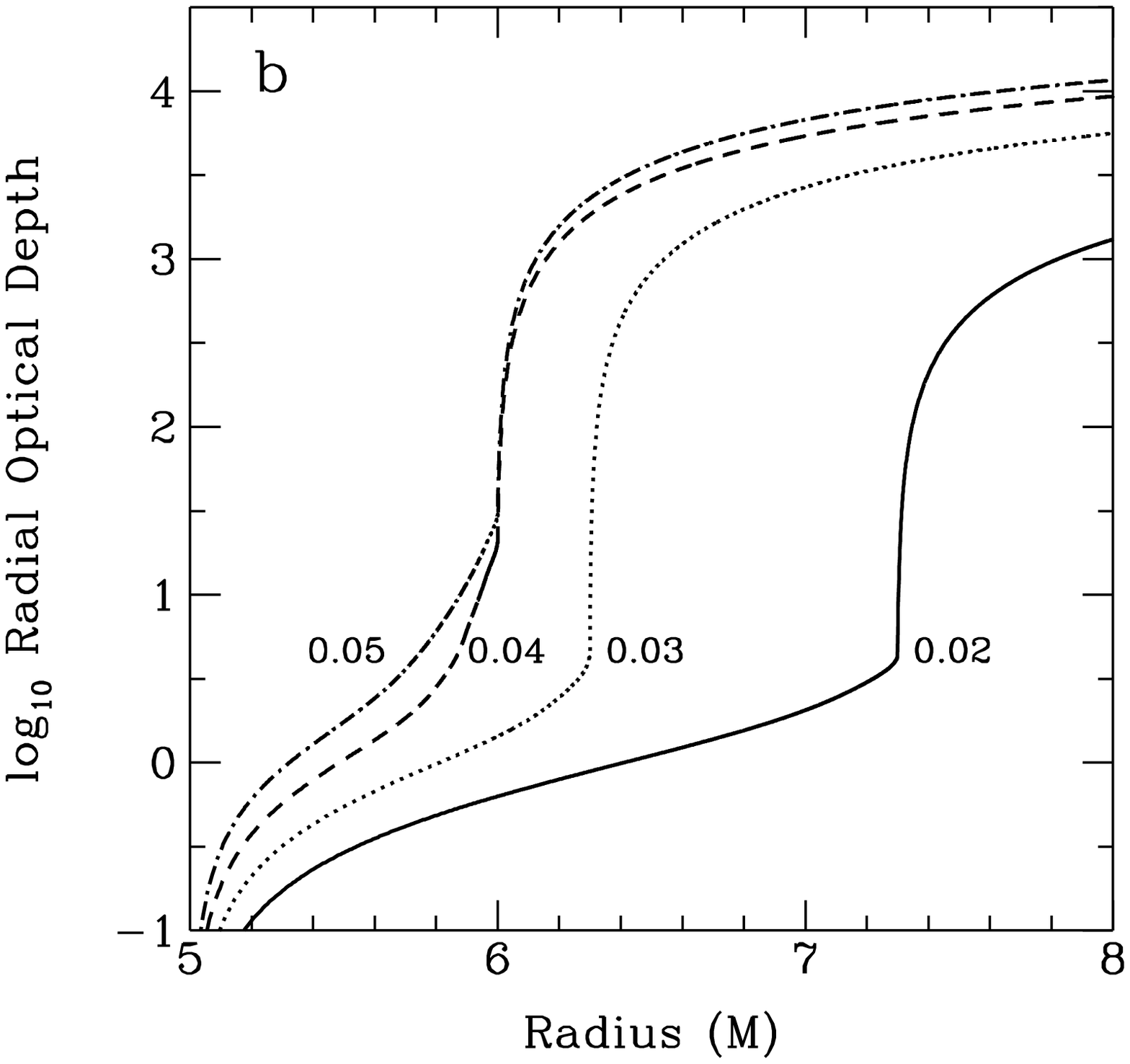,height=3.4truein}}
 }
 \vglue-1.0truecm
 \caption{
 Results of fully general relativistic
numerical computations of the gas dynamics
and radiation transport in the inner disk
in the sonic-point model, for a neutron
star of radius $5M$.
 (a)~Inward radial velocity $v^{\hat r}$ of
the gas in the disk measured by a local
static observer, as a function of the
Boyer-Lindquist radial coordinate expressed
in units of the stellar mass $M$.
 (b)~Radial optical depth from the
stellar surface through the disk flow to
the radius shown on the horizontal axis.
 Each curve is labeled with the assumed
accretion rate $\dot M$ measured in units
of the accretion rate $\dot M_E$ that would
produce an accretion luminosity at infinity
equal to the Eddington critical luminosity.
 From MLP.}
 \end{figure*}

Gas streaming inward from clumps orbiting
near the sonic radius along trajectories
with the tight spiral shape shown in
Figure~3a generates the more open spiral
density pattern shown in Figure~3b.
Collision of the denser gas from the clumps
with the stellar surface creates beams of
brighter X-ray emission, like the beam
indicated by the white dashed lines in
Figure~3b. These beams move around the
star's equator, generating a quasi-periodic
brightness oscillation with frequency
$\nu_{\rm Ks}$. The lower-frequency QPO is
generated by weak X-ray beams produced by
funneling of part of the accretion flow
near the star by the star's weak magnetic
field. These beams rotate {\em with the
star\/} and modulate the radiation drag
acting on the gas at the sonic radius,
modulating the inward mass flux and the
luminosity at the sonic-point beat
frequency $\nu_{\rm Bs}$ (\hbox{$\nu_{\rm
Ks}-\nu_{\rm spin}$} or \hbox{$\nu_{\rm
Ks}-2\nu_{\rm spin}$}).

 \begin{figure*}[t!] 
 \label{spirals}
 \vglue-2.2truecm
 \centerline{\hskip0.8truecm
\psfig{file=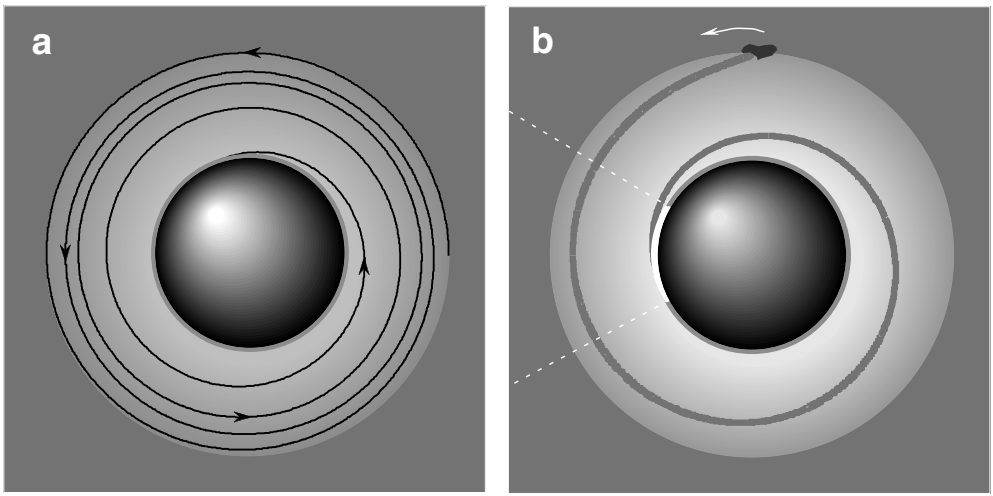,angle=90,height=12.0truecm}}
 \vglue-3.2truecm
 \caption{View of the neutron star and disk
along the rotation axis of the disk,
which is rotating counterclockwise in this
view.
  (a)~Spiral trajectory followed by a
single element of gas as it falls
supersonically from the sonic radius to the
stellar surface.
  (b)~Spiral pattern of higher gas density
formed by gas streaming inward along spiral
trajectories with the shape shown in (a),
from a clump orbiting near the sonic radius.
 The spiral trajectory and density pattern
are from fully general relativistic
calculations (see MLP).}
 \end{figure*}

The sonic-point model is consistent with
the accretion rates, stellar magnetic
fields, and scattering optical depths
inferred previously from \exosat\ and
\ginga\ observations of the atoll and Z
sources and accounts for the main features
of the kilohertz QPOs, including their high
and variable frequencies, their high
amplitudes and coherences, and the common
occurrence of kilohertz QPOs in pairs (see
MLP). It also explains naturally why the
frequency separation between the
frequencies of a kilohertz QPO pair is
nearly constant and equal to the burst
oscillation frequency or half this
frequency. Finally, the sonic-point model
can account for the similar frequency ranges
of the kilohertz QPOs in sources with very
different accretion rates and magnetic
fields.

\section{Implications for Neutron Stars}

The frequency of the higher-frequency QPO
in a kilohertz QPO pair is almost certainly
the orbital frequency of gas in Keplerian
orbit around the star. If so, the high
frequency and coherence these QPOs allow
one to derive tight upper bounds on the
masses and radii of the neutron stars that
produce such QPOs and significant
constraints on the equation of state of
neutron star matter.

To see how such bounds can be constructed,
suppose first that the star is not rotating
and assume that, for the star in question,
$\nu_{\rm QPO2}^\ast$---the highest
observed value of the frequency of the
higher-frequency (Keplerian-frequency) QPO
in the kilohertz QPO pair---is 1220~Hz
(this is the highest QPO frequency detected
so far in any source; see MLP). Obviously,
the radius $R_{\rm orb}$ of the orbit of
the gas producing the QPO must be greater
than the radius $R$ of the star. This means
that the star's representative point in the
$R$,$M$ plane must lie to the left of the
cubic curve $M^\nonrot(R_{\rm orb})$ (the
dashed curve shown in Figure~4a) which
relates the star's mass to the radius of
orbits with frequency 1220~Hz. In order to
produce a wave train with tens of
oscillations, $R_{\rm orb}$ must also be
greater than the radius $R_{\rm ms}$ of the
innermost stable circular orbit, so the
actual radius of the orbit must lie on the
portion of the $M^\nonrot(R_{\rm
orb},\nu_{\rm QPO2}^\ast)$ curve that lies
below its intersection with the diagonal
line $M^\nonrot(R_{\rm ms})$ (the dotted
line shown in Figure~4a) which relates the
star's mass to $R_{\rm ms}$. As Figure~4a
shows, this requirement bounds the mass and
radius of the star from above. For $\nu_{\rm
QPO2}^\ast = 1220~\Hz$, the representative
point of the star must lie in the pie-slice
shaped region enclosed by the solid lines in
Figure~4a. Thus, the mass and radius of a
nonrotating star with this QPO frequency
would have to be less than $1.8\,\msun$ and
16.0~km, respectively. Figure~4b compares
the mass-radius relations for nonrotating
stars given by five equations of state with
the region of the radius-mass plane allowed
for three values of $\nu_{\rm QPO2}^\ast$.

 \begin{figure*}
 \vglue-1.5truecm
\centerline{\psfig{file=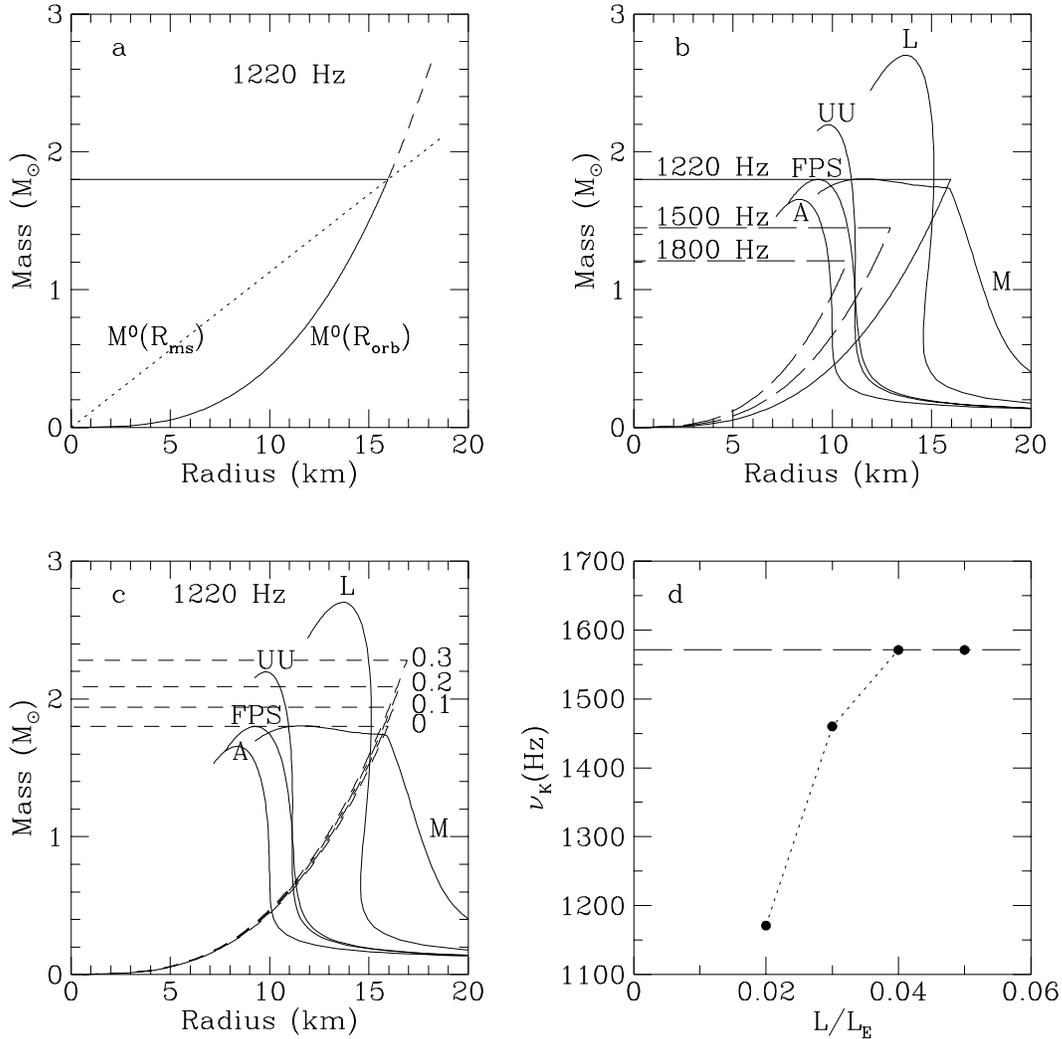,height=6truein,width=6truein}}
 \vglue-0.5truecm
 \caption{
 (a)~Radius-mass plane, showing how to
construct the region allowed for a
nonrotating neutron star with $\nu_{\rm
QPO2}^\ast = 1220$~Hz (see text).
  (b)~Comparison of the mass-radius
relations for nonrotating neutron stars
given by five representative equations of
state with the regions of the mass-radius
plane allowed for nonrotating stars with
three different QPO frequencies. The light
solid curves show the mass-radius relations
given by equations of state A, FPS, UU, L,
and M.
  (c)~Regions allowed for rotating neutron
stars with various values of $j$ and
$\nu_{\rm QPO2}^\ast = 1220$~Hz, when
first-order effects of the stellar spin
are included.
  (d)~Illustrative Keplerian QPO frequency
as a function of accretion luminosity given
by the general relativistic calculations
described in \S~2.
 For details, see MLP.
 }
 \end{figure*}

The allowed region of the $R$,$M$ plane is
affected by rotation (see MLP). The
parameter that characterizes the importance
of rotational effects is the dimensionless
quantity \hbox{$j \equiv cJ/GM^2$}, where
$J$ and $M$ are the angular momentum and
gravitational mass of the star. For the
spin frequencies $\sim$300~\Hz inferred in
the kilohertz QPO sources, $j$ is $\sim 0.1
\dash 0.3$, depending on the equation of
state. Figure~1c illustrates the effects of
slow stellar rotation on the allowed region
of the $R$,$M$ plane. The region allowed
for a slowly rotating star is always larger
than the region allowed for a nonrotating
star, regardless of the equation of state.
However, the region allowed for a rapidly
rotating star can be smaller than that for
the corresponding nonrotating star (Miller,
Lamb, \& Cook 1998).

If the frequency of a kilohertz QPO can be
established as the orbital frequency of gas
at the innermost stable circular orbit,
this would be an important step forward in
our understanding of strong-field gravity
and the properties of dense matter, because
it would confirm one of the key predictions
of general relativity in the strong-field
regime and fix the mass of the neutron star
in that source, for each assumed equation
of state. Probably the most convincing
signature would be a fairly coherent
kilohertz QPO with a frequency that
reproducibly increases steeply with
increasing accretion rate but then becomes
constant and remains nearly constant as the
accretion rate increases further. This
behavior emerges naturally from general
relativistic calculations of the gas
dynamics and radiation transport in the
sonic-point model (see Fig.~4d). The
constant frequency should always be the
same in a given source.

\medskip

This work was supported in part by NSF grant
AST~96-18524, NASA grant NAG~5-2925, and
NASA RXTE grants at the University of
Illinois, and by NASA grant NAG~5-2868 at
the University of Chicago.

\section*{References}

 \re
  Klein, R.\,I., Jernigan, J.\,G., Arons,
J., Morgan, E.\,H., \& Zhang, W. 1996, ApJ,
469, L119
 \re
 Lamb, F.~K. 1991, in Neutron Stars: Theory
and Observation, ed. J. Ventura \& D.
Pines, (Dordrecht: Kluwer), 445
  \re
  M\'endez, M., et al. 1997, ApJ, in press
  (preprint astro-ph/9712085)
 \re
 Miller, M.\,C., \& Lamb, F.\,K. 1996, ApJ,
470, 1033
 \re
 Miller, M.~C., Lamb, F.~K., \& Cook, G.
 1998, in preparation
 \re
 Miller, M.\,C., Lamb, F.\,K., \& Psaltis,
D. 1998, ApJ, in press (MLP)
 \re
 Smith, D.\,A., Morgan, E.\,H., \& Bradt,
H. 1997, ApJ, 479, L137
 \re
 Strohmayer, T.\,E., Swank, J.\,H., Zhang,
W. 1998, in The Active X-Ray Sky, ed. L.
Scarsi, H. Bradt, P. Giommi, \& F. Fiore,
Nuclear Phys. B Proc. Suppl., in press
(astro-ph/9801219)
 \re
 Strohmayer, T., Zhang, W., Swank, J.\,H.
1997, ApJ, 487, L77
 \re
 \re
 Strohmayer, T. 1997, talk presented at the
1997 HEAD Meeting, Estes Park, Colorado
 \re
 Titarchuk, L., \& Muslimov, A. 1997, A\&A,
323, L5
 \re
 van der Klis, M. 1997, in The Many Faces
of Neutron Stars, Proc. NATO ASI, Lipari,
Italy (Dordrecht: Kluwer), in press
(astro-ph/9710016)
 \re
 van der Klis, M., Wijnands, R., Horne, K.,
\& Chen, W. 1997, ApJ, 481, L97
 \re
 Wijnands, R.\,A.\,D., \& van der Klis, M.
1997, ApJ, 482, L65
 \re
 Zhang, W., Lapidus, I., Swank, J.\,H.,
White, N.~E., \& Titarchuk, L. 1997, IAU
Circ. 6541

\vspace{1pc}

\end{document}